\documentclass{cmip-l}
\usepackage{amssymb,epsf}
\newcommand{\be}{\begin{equation}}
\newcommand{\ee}{\end{equation}}
\newcommand{\bea}{\begin{eqnarray}}
\newcommand{\eea}{\end{eqnarray}}
\newcommand{\beas}{\begin{eqnarray*}}
\newcommand{\eeas}{\end{eqnarray*}}

\def\vertol{\;\raisebox{-1mm}{\epsfysize=4mm\epsfbox{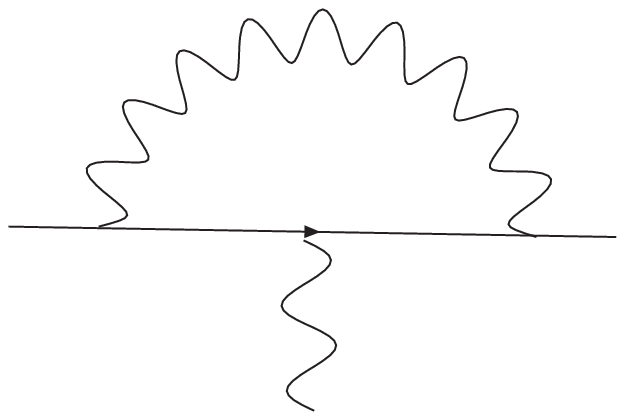}}\;}

\def\verttl{\;\raisebox{-4mm}{\epsfysize=10mm\epsfbox{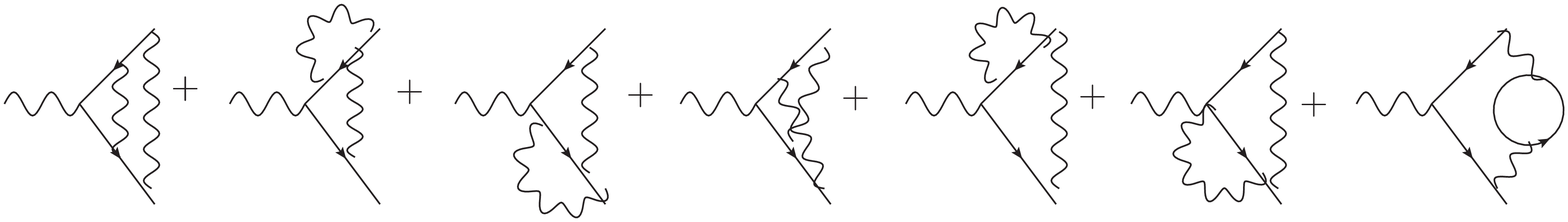}}\;}
\def\phol{\;\raisebox{-1mm}{\epsfysize=4mm\epsfbox{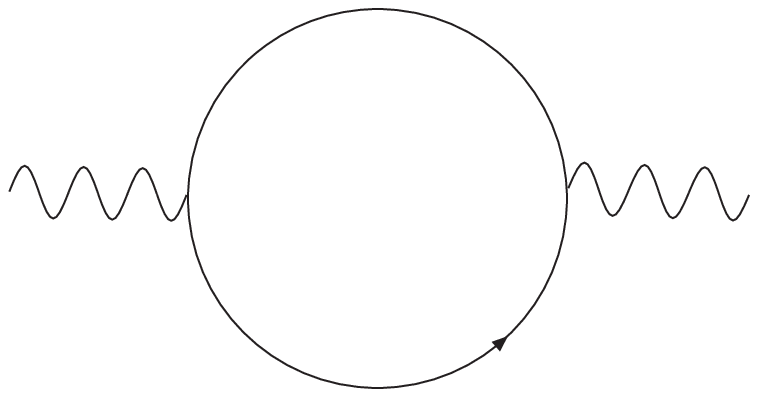}}\;}
\def\phtl{\;\raisebox{-3mm}{\epsfysize=8mm\epsfbox{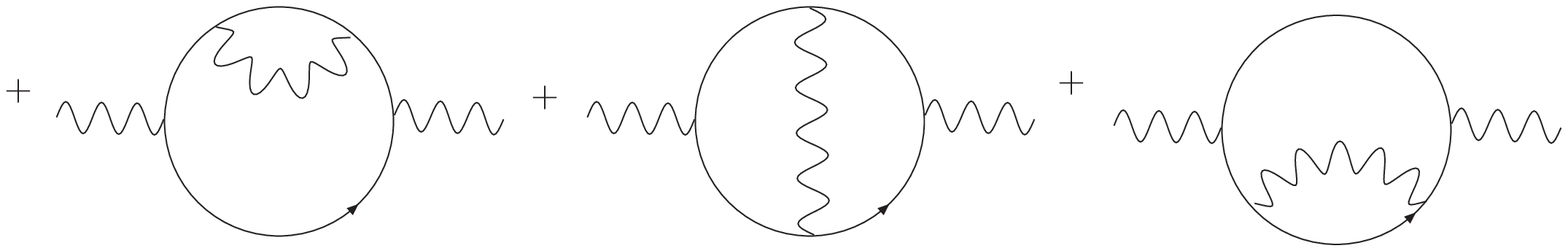}}\;}
\def\feol{\;\raisebox{-1mm}{\epsfysize=4mm\epsfbox{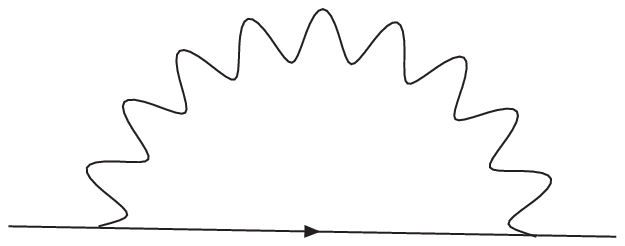}}\;}
\def\feolup{\;\raisebox{1mm}{\epsfysize=4mm\epsfbox{fpol.eps}}\;}
\def\feolol{\;\raisebox{-2mm}{\epsfysize=6mm\epsfbox{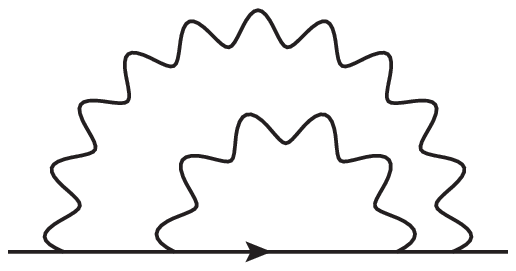}}\;}
\def\feololsmall{\;\raisebox{1mm}{\epsfysize=4mm\epsfbox{feolol.eps}}\;}
\def\feolcm{\;\raisebox{-2mm}{\epsfysize=6mm\epsfbox{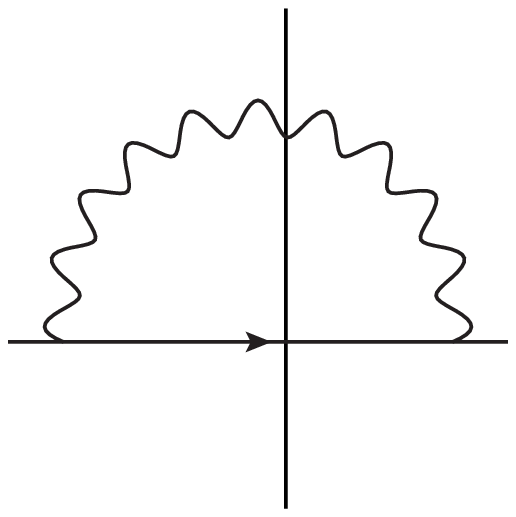}}\;}
\def\feololcm{\;\raisebox{-2mm}{\epsfysize=6mm\epsfbox{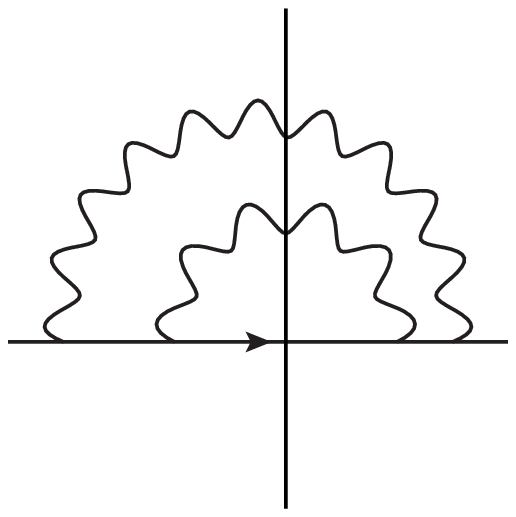}}\;}
\def\feololcl{\;\raisebox{-2mm}{\epsfysize=6mm\epsfbox{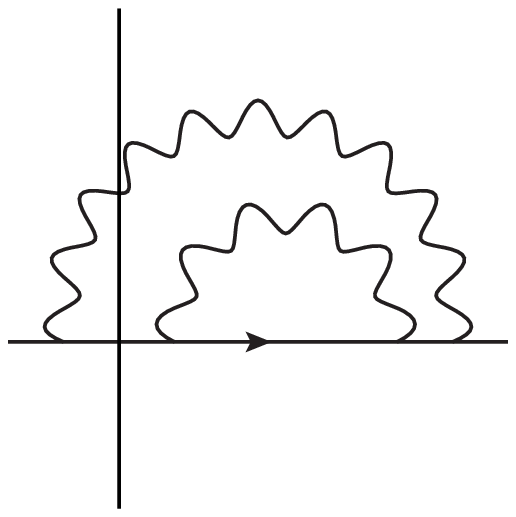}}\;}
\def\feololcr{\;\raisebox{-2mm}{\epsfysize=6mm\epsfbox{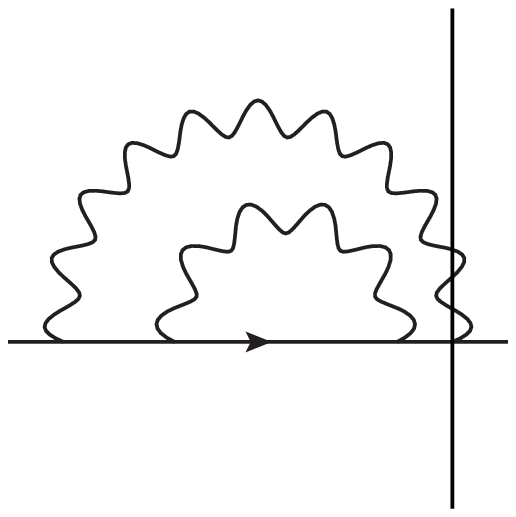}}\;}
\def\fetl{\;\raisebox{-3mm}{\epsfysize=8mm\epsfbox{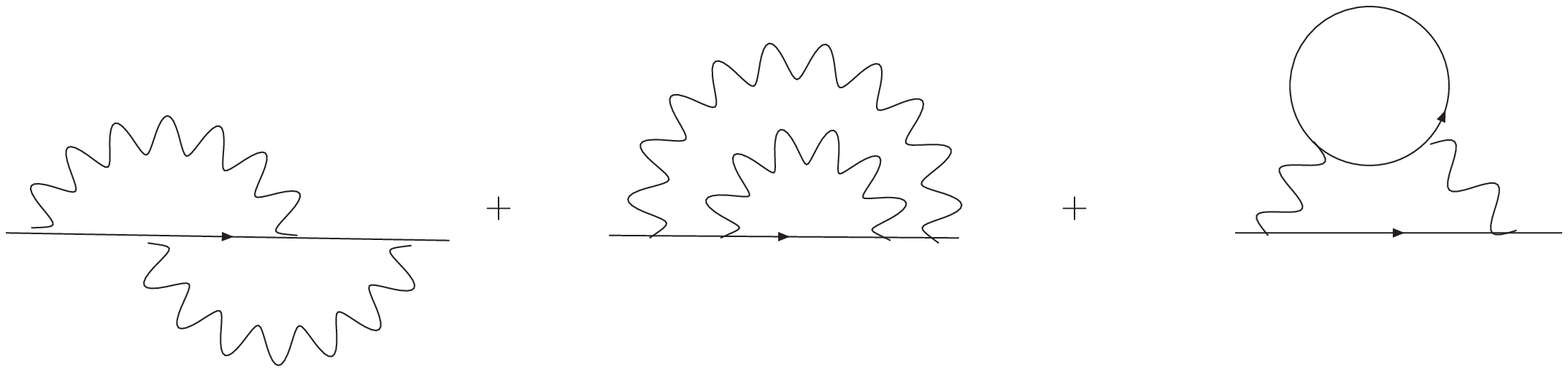}}\;}

\def\qedtlone{\;\raisebox{-3mm}{\epsfysize=7mm\epsfbox{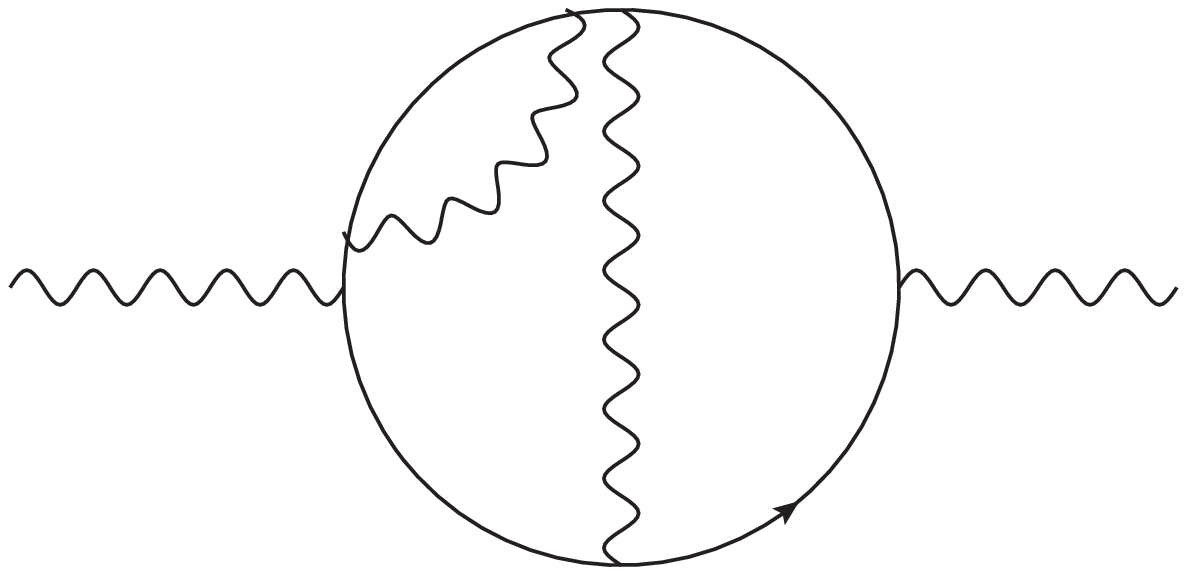}}\;}
\def\qedtlonetree{\;\raisebox{-8mm}{\epsfysize=20mm\epsfbox{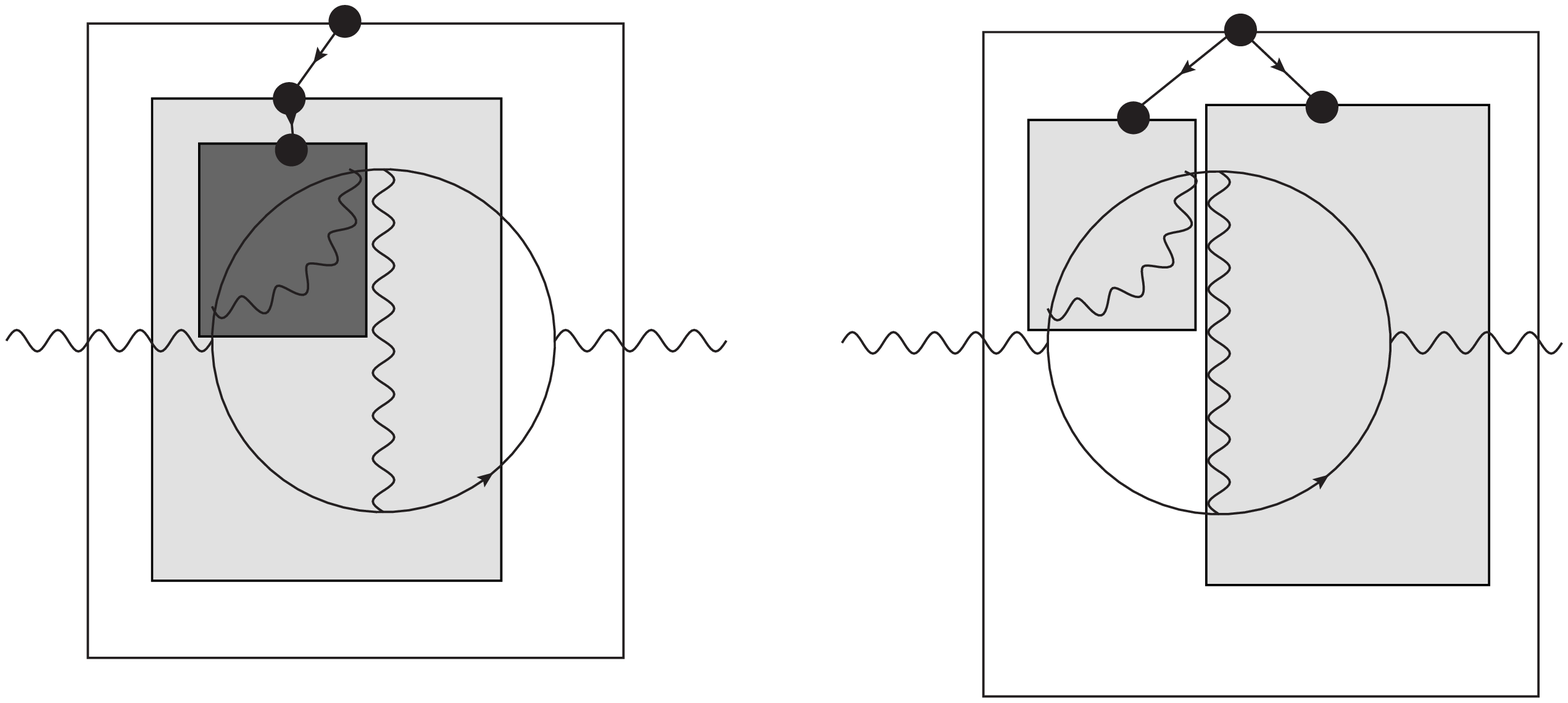}}\;}
\def\qedtltwo{\;\raisebox{-3mm}{\epsfysize=7mm\epsfbox{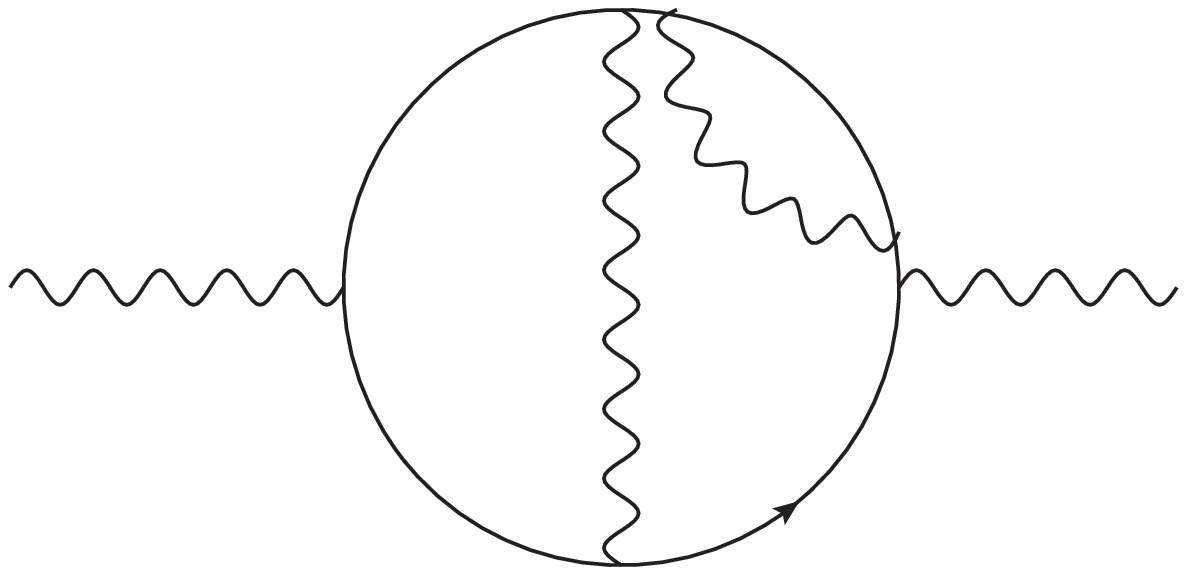}}\;}
\def\qedtlthree{\;\raisebox{-3mm}{\epsfysize=7mm\epsfbox{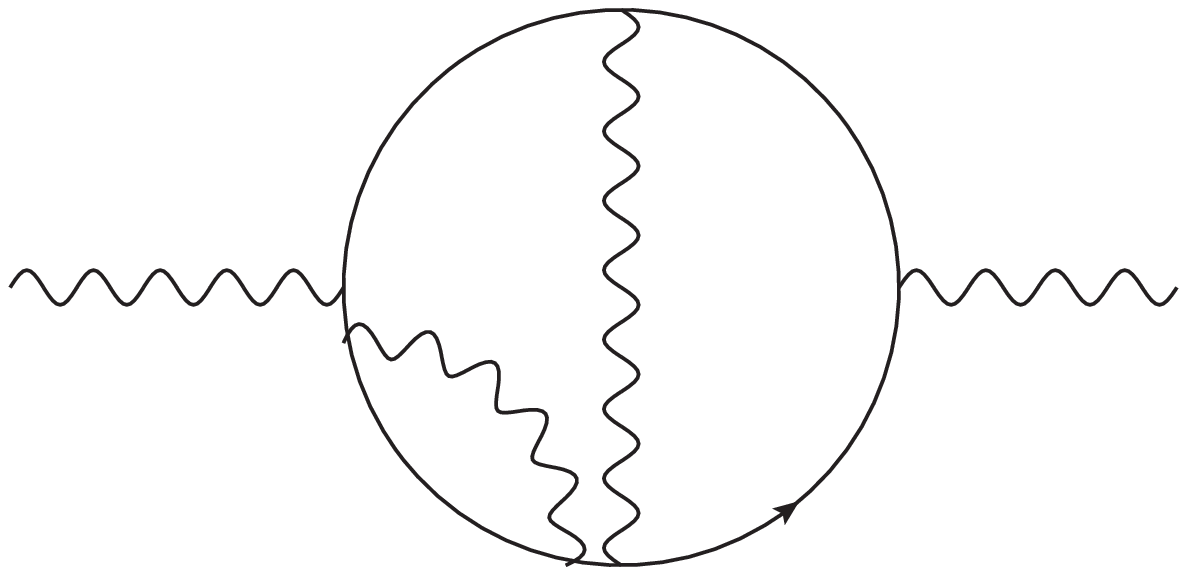}}\;}
\def\qedtlfour{\;\raisebox{-3mm}{\epsfysize=7mm\epsfbox{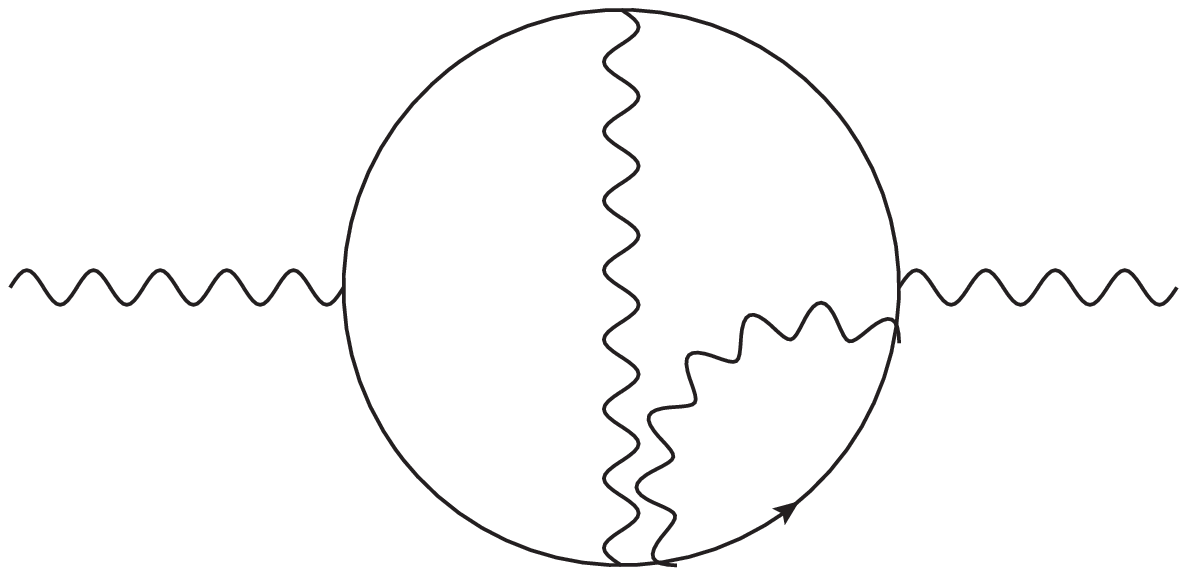}}\;}
\def\qedtwlone{\;\raisebox{-2mm}{\epsfysize=6mm\epsfbox{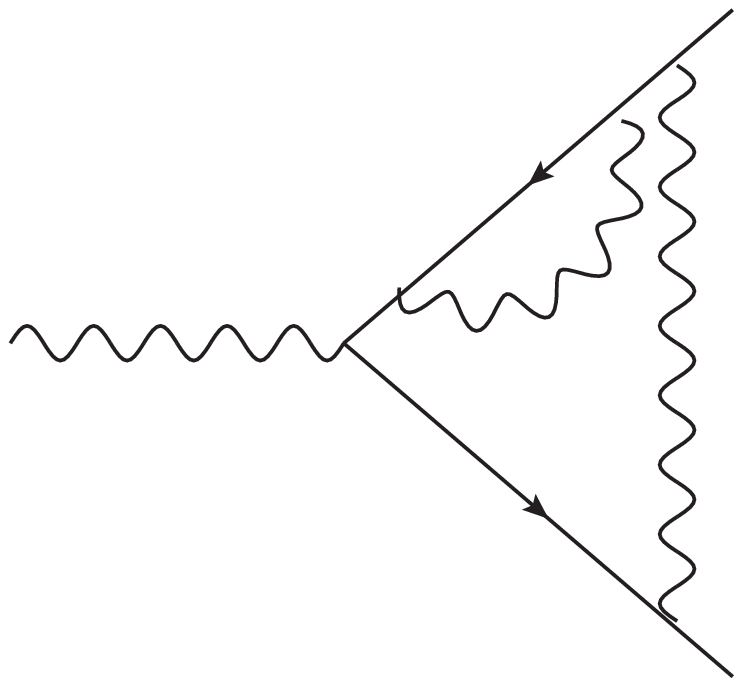}}\;}
\def\qedtwltwo{\;\raisebox{-2mm}{\epsfysize=6mm\epsfbox{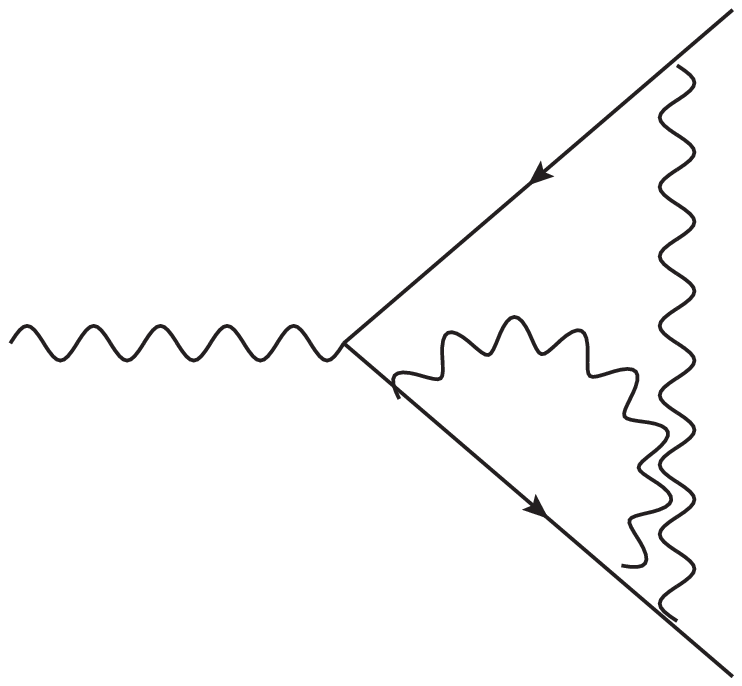}}\;}
\def\qedtl{\;\raisebox{-2mm}{\epsfysize=6mm\epsfbox{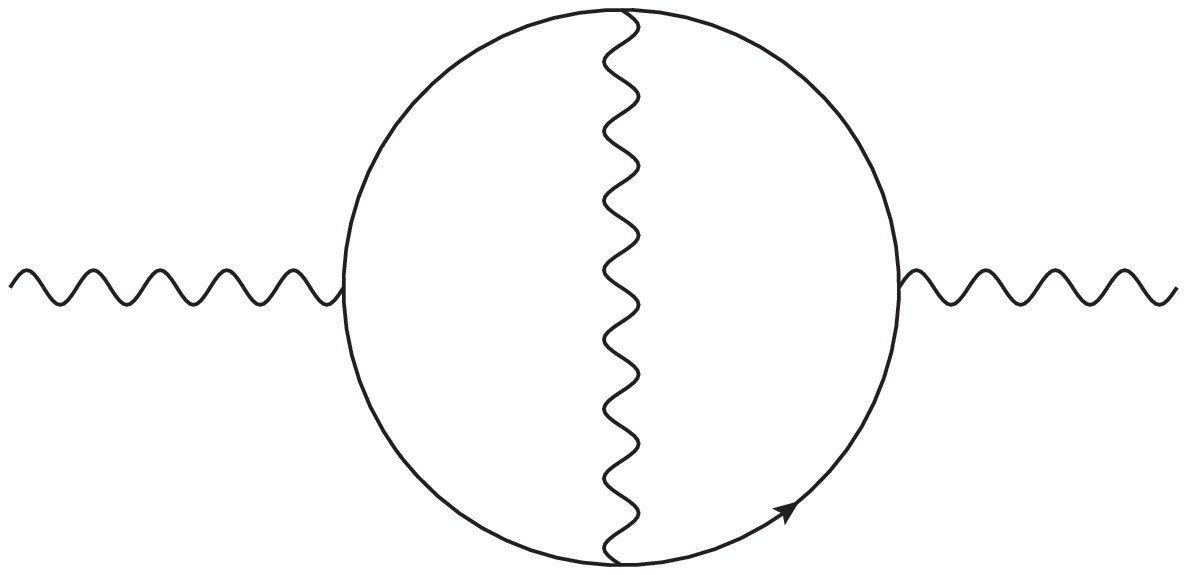}}\;}
\def\qedtlf{\;\raisebox{-2mm}{\epsfysize=6mm\epsfbox{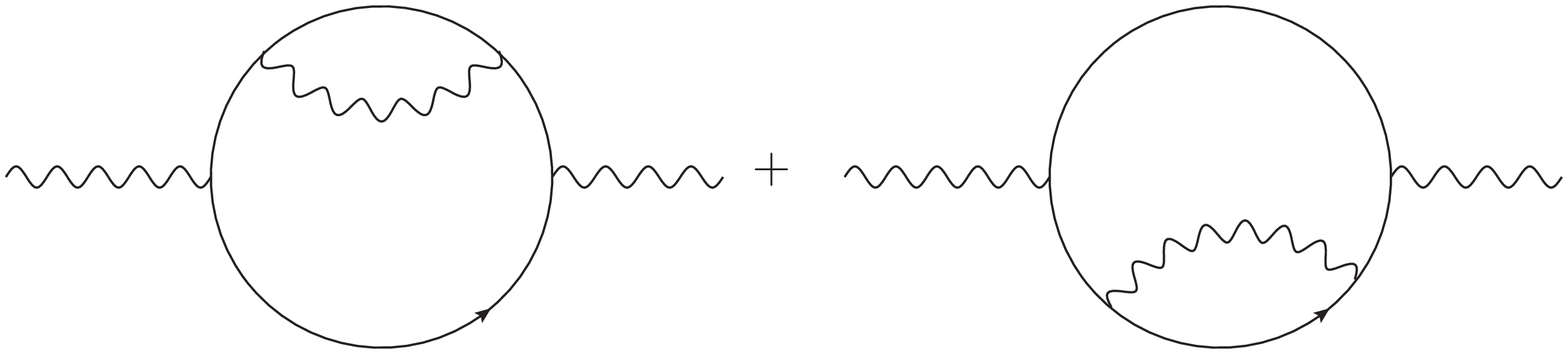}}\;}

\def\One{\mathbb{I}}

\title{Algebra for quantum fields}
\author{Dirk Kreimer}
\address{kreimer@ihes.fr, IHES, 35 rte. de Chartres, 91440 Bures-sur-Yvette, France (http://\ www.ihes.fr)
and Boston U.\ (http://math.bu.edu)}
\thanks{${}^*$Contributed to the proceedings for the BU conference. Work supported in parts by grant NSF-DMS/0603781. Author supported by CNRS}
\begin{document}
\begin{abstract}
We give an account of the current state of the approch to quantum field theory via 
Hopf algebras and Hochschild cohomology.
We emphasize the versatility and mathematical foundation of this algebraic structure, 
and collect algebraic structures here in one place which are either scattered over the literature, or only implicit in previous writings.
In particular we  
point out  mathematical structures which can be helpful to farther develop our mathematical understanding of quantum fields.
\end{abstract}
\maketitle
\section*{Introduction}
\subsection*{Acknowledgments}
It is a pleasure to thank my former students Christoph Bergbauer, Kurusch Ebrahimi-Fard and Karen Yeats for discussions and advice.
\subsection*{The reason for QFT} What is quantum field theory (QFT) about? For that matter, 
what is quantum physics about? The answer, given with the necessary grain of pragmatism,
is simple: sum over all histories connecting a chosen initial state with a particular final state. Square that complex-valued sum.

Various attempts had been made to make this paradigm precise: the desired ``sum over histories'' has not yet reached its final form though 
- mathematicians are still,
and rightfully so, baffled by QFT. 

Physicists have created myriads of examples meanwhile where one can stretch its mind and flex its muscles on what is usuaully called the path integral. 
Many results point to rather fascinating structures, 
carefully formulated in a self-consistent way. A mathematical definition of said path integral beyond perturbaton theory is lacking though, 
leaving the author with considerable unease.

Early on, it was recognized that the desired Green functions in field theory are constrained by quantum equations of motion, 
the Dyson--Schwinger equations. The latter suffer from short-distance singularities, leading to the need for renormalization. 

It took us physicists a while to learn how to handle this routinely, and in a mathematical consistent way.
Progress was made by elaborate attempts at low orders of perturbation theory, 
and the above equations of motions took a backseat in contemporary physics, while formal approaches 
starting from the functional integral (constructive methods)
typically failed to come to conclusions for renormalizable theories, which remain theories of prime interest.

Meanwhile, perturbative renormalization was embarrassingly effective in describing reality, and kept QFT, understood as an expansion in Feynman graphs,
 in its role as the best-tested and most-used 
workhorse in the stables of theoretical physics. It is commonly denied the status of being a theory these days though, 
as at the moment of writing it is not
yet defined in mathematically satisfying terms.

It is a personal belief of the author that this is not testimony for bigger (read extended) things hiding behind local quantum fields,
but rather testimony to the subtlety by which nature hides its concepts.

On an optimistic nore, I indeed believe that the clear mathematical understanding we have now 
of the practice of perturbative renormalization paves the way 
for a mathematical consistent approach to QFT which bridges the gap between what practitioners of QFT have learned, 
and what is respectable mathematics.

The approach exhibited in the following is based solely on representation theory of the Poincar\'e group 
and the requirement that interactions are local.

Before we start, let me emphasize that here is not the place to comment in detail on progress with analytic aspects. 

Still, let me mention two encouraging developments: 
non-perturbative aspects can now be studied using Dyson--Schwinger equations in a much more effective manner 
\cite{KY1,KY3,KY4,Bellon1,Bellon2}, and the relation to periods and motivic theory, became a (little!)  bit clearer 
in collaboration with Esnault and Bloch \cite{BEK,BlKr}.

In particular, Feynman integrals are periods \cite{BelBros,Brow,BEK,BlKr} (considered as a function of masses and external momenta, they are periods when those
parameters take rational values \cite{WeinB}, though a much better argument should be made for the Taylor coefficients in the expansion in such variables).

Better still, these periods are interesting: in a suitable parametric representation based on edge variables $A_e$ for edges $e$ \cite{BEK,BlJap,BlKr}, 
they appear as periods of the mixed Hodge structure on
the middle-dimensional cohomology 
$$ H^{2m-1}(P \backslash Y_\Gamma,B \backslash B\cap Y_\Gamma)$$ 
constructed from blow-ups $P\to\ldots\to \mathbb{P}^{2m-1}$ which separate the boundary of the chain of integration (contained in $\Delta=
\cup_{e\in E(\Gamma)}{A_e=0}$) from the
singularities of the graph hypersurface $X_\Gamma$, with $Y_\Gamma$ the strict transform of $X_\Gamma$ and $B$ the inverse image of $\Delta$.

For example, the complete graph on four vertices (a contribution to the vertex function of $\phi^4$ theory) has a period $6\zeta(3)$ contained in 
$\zeta(3)\mathbb{Q}$, \cite{BEK,BlJap}. Such results have been recently extended by Dzimitry Doryn \cite{DorynThesis}.

Now back to the underlying algebraic structures. Lets first get edges and vertices for our graphs.

\section{Free QFT, interacting QFT}
Classical geometry does not rule the day when it comes to quantum fields. 
Much to the contrary, the often beautiful classical geometry of fields, gauge fields in particular, 
must emerge as a classical limit of quantum field theory.  Hence we speak about QFT without taking recourse to classical fields. 
We  ignore the geometry of the classical spacetime manifold over which we want to construct QFT, 
and just memorize that it has a four-dimensional tangential and cotangential space locally, 
isomorphic to flat Minkowski space.  It is over such local fibres that one formulates QFT.

Our first concern is to understand the elementary amplitudes 
which we use to describe the observable physics which results from quantum field theory.

They come in two garden varieties: amplitudes for propagation, and amplitudes for scattering.
The former are provided by free quantum field theory: free propagators, 
in momentum space, are obtained as the inverse of the free covariant wave equations. 
Hence, for Minkowski space, its Wigner's representation theory of the Lorentz and Poincar\'e groups which rules the day, 
providing us with free propagators for massless and massive bosons and fermions.

The latter, amplitudes for scattering, are again provided by representation theory of the Poincar\'e group, augmented by 
the requirement for locality.

Let us look at a simple example to see how this comes about. Assume we take from free quantum fields 
the covariances for a free propagating electron, positron and photon. Assume we want to couple those in a local interaction.
Representation theory tells us that this interaction will have to transform as a 
Lorentz vector $v_\mu$, coupling the spin-one photon to a spin-1/2
electron and positron. Also, knowing the scaling weights of free photons, electrons and positrons as determined 
from the accompanying free field monomials, such a vertex must have zero scaling weight itself,
as the scaling weights of those monomials add up to the dimension of spacetime.
Indeed,
\bea [ \bar{\psi}\partial\!\!\!/ \psi]  =  4 & \Rightarrow & [\bar{\psi}]=[\psi]=3/2,\\
\frac{1}{4}[F^2]  =  4 & \Rightarrow & [A]=1,\eea
\be [v_\mu \bar{\psi} A_\mu \psi]   =  4  \Rightarrow  [ v_\mu]=0.\ee

So what would be the Feynman rule, in momentum space say, for such an amplitude?
If the electron has momentum $p_1$, the positron momentum $p_2$, and the photon momentum $q=-p_1-p_2$,
the vertex can be a linear combination of twelve invariants \be v_\mu=c_1\gamma_\mu+c_2\frac{q_\mu q\!\!\!/}{q^2}+\cdots.\ee

But if we have to have a local theory, any graph for a quantum correction for the unknown vertex built 
from that unknown vertex and the known propagators 
will be, by a simple powercounting exercise, -we know the scaling weight of our unknown vertex at least-, logaritmic divergent.

If we are to absorb this logarithmic divergence by a local counterterm, this gives us information on the desired Feynman rule. Let us  work it out.
To keep the example simple, let us assume we suspect that the vertex is of the form
\be v_\mu=v_\mu(q)=c_1\gamma_\mu+c_2\frac{q_\mu q\!\!\!/}{q^2}.\ee
Let us consider the one-loop 1PI graph -the lowest order quantum correction- to find the sought after Feynman rule.

With three vertices in the graph we have $2^3=8$ integrals to do which appear in the limit
\beas \lim_{\Lambda\to+\infty}\frac{1}{\ln\Lambda}\Phi\left(\vertol\right) & \sim &  
\lim_{\Lambda\to+\infty}\frac{1}{\ln\Lambda}
\int_{-\Lambda}^{\Lambda} v_\alpha(k)\frac{1}{k\!\!\!/}v_\mu(k)\frac{1}{k\!\!\!/}v_\beta(k) D_{\alpha\beta}((k+q)^2)d^4k\nonumber\\ & = & 
f(c_1,c_2)\gamma_\mu.\eeas
In a local theory, the coefficient of $\ln \Lambda$ from the integral must be proportional to the desired vertex. Hence, dividing and taking the limit, 
we
confirm that  the term $\sim q_\mu q\!\!\!/$ vanishes like $1/\ln \Lambda$ in all eight terms.
We hence conclude $c_2=0$, and this gives a good idea how locality is needed for quantum field theory to stabilize 
at dintinguished Feynman rules in a self-similar manner. Similarly, if we had done the example with the full $12^3$ terms of the full vertex,
as it must be for a renormalizable theory.

We also conclude that the price for Feynman rules determined by locality is that we indeed pick up local short-distance singularities. That leaves us the freedom to set a scale,
which is no big surprise: looking only at quantum fields for a typical fibre -the cotangential space-, we hence miss the only parameter around to set a scale: 
the curvature of the underlying manifold. 
The extension of such local notions to the whole manifold awaits understanding of quantum gravity. This might well start from understanding how
gravity with its peculiar powercounting behaves as a Hopf algebra \cite{grav}.

Let us now proceed to see what comes with those edges and vertices as prescribed above - graphs, obviously.
\section{1PI graphs, Hopf algebras}
Having hence elementary scattering and propagation amplitudes available, we can set up a quantum theory:
we define incoming and outgoing asymptotic states, and sum over all unobserved intermediate states. This is standard material for a physicist, 
and we leave it to the reader to acquaint himself with the necessary details on the LSZ formalism and other such aspects \cite{Wightm,BS,BjDr,ItzZ}.

While many textbooks on contemporary physics proceed 
using the path integral to define Green functions for amplitudes, for connected amplitudes and for 1PI amplitudes, 
we emphasize that these Green functions can be given mathematical precise meaning through the study of the Hopf algebra structure underlying
the graphs constructed from the representation theory mentioned above.\footnote{This might implicitly define the path integral, which has to be seen
in future work. Too often in the authors opinion, the path integral is in the context of quantum field theory only a reparametrization 
of our lack of understanding, giving undue prominence to the classical Lagrangian.}

So having Feynman rules for edges and vertices the above  gives us Feynman rules for n-PI graphs, 
graphs which do not disconnect upon removal of any n internal edges. 
Amplitudes for connected graphs are obtained from 1-PI Green 
functions by connecting them via free covariances, and disconnected graphs finally by exponentiation.
Its for 1-PI graphs that the underlying algebraic structure of field theory becomes fully visible.

The basic such algebraic structure then at our disposal are:\\ 
i) the Hopf- and Lie algebras coming with such graphs,\cite{K,CK,RHI,RHII}\\ 
ii) the correponding Hochschild cohomology and the sub-Hopf algebras generated by the grading,\cite{BergbKr,Foissy}\\ 
iii) the co-ideals corresponding to symmetries in the Lagrangian,\cite{anatomy,Walter}\\ 
iv) the coradical filtration and Dynkin operators governing the renormalization group and leading log expansion,\cite{BroadK,KY2,Patras}\\ 
v) the semi-direct product structure between superficially convergent and divergent graphs, \cite{RHI}\\  
vi) and finally the core Hopf algebra \cite{BlKr,KrCMI}, suggesting co-ideals leading to recursions \`a la 
BCFW, showing that loops and legs speak to each other in many ways: 
it is indeed the hope of the author that the rather disparate structures  we observe in experience with multi-loop vs multi-leg expansions 
combine finally in a common mathematical framework \cite{WvSDK}. 

We omitted in this list
Rota--Baxter algebras \cite{Kurusch}, 
which are useful for MS schemes but less so  in renormalization schemes based on on-shell or momentum space subtractions. The reader can find
detailed study of Rota--Baxter algebras in the above-cited work of Ebrahimi-Fard and Manchon, while the use of momentum space subtractions 
was exhibited recently beyond perturbation theory in \cite{KY3}.
We also omit the algebraic structure
of field theory in coordinate space, see \cite{BergbDipl,BergbKEG,BergbDiss} for a clarification how to connect it with the approach described here. 

In this contribution, we will mainly review combinatorial and algebraic aspects developed in recent years. 
We include a few results only implicit in published work so far.
A summary of analytic and algebro-geometric achievements has to be given elsewhere.

Let us now illustrate these algebraic structures. For that we
strengthen our muscle on quantum electrodynamics  (QED)  graphs for the vertex, fermion- and photon-selfenergy, up to two loops each.
Here they are:
\bea 
c_1^{\bar{\psi}A\!\!\!/\psi} & = & \vertol,\\ 
c_2^{\bar{\psi}A\!\!\!/\psi} & = & \verttl,\\
c_1^{\bar{\psi}\psi} & = & \feol,\\
c_2^{\bar{\psi}\psi} & = & \fetl,\\
c_1^{ \frac{1}{4}F^2 } & = & \phol,\\
c_2^{ \frac{1}{4}F^2 } & = & \phtl.
\eea 
\subsection{The Hopf algebra}
We define a family of Hopf algebras $\mathcal{H}$. 
Each Hopf algebra $H\in \mathcal{H}$ is generated by generators given by 1-PI graphs 
and its algebra structure is given as the free commutative $\mathbb{Q}$-algebra 
over those generators, with the empty graph furnishing the unit $\One$.

For a graph $\Gamma$, we let $\Gamma^{[0]}$ be the set of its vertices, $\Gamma^{[1]}_{\mathrm{int}}$ the set of its internal edges,
and $\Gamma^{[1]}_{\mathrm{ext}}$ be the set of its external edges. Each edge is assigned an arbitrary orientation 
(all physics is independent of that choice), so that we can speak of a source $s(e)$ and target $t(e)$ for an edge $e$. 
For each internal edge $e\in\Gamma^{[1]}_{\mathrm{int}}$, $s(e)\in \Gamma^{[0]}$ and $t(e)\in \Gamma^{[0]}$. We do not require
that $s(e)\not= t(e)$. For each $e\in \Gamma^{[1]}_{\mathrm{ext}}$, $t(e)\in \Gamma^{[0]}$ but $s(e)\not\in \Gamma^{[0]}$.

To each internal edge $e$ we assign a weight $w(e)$, to each vertex $v$ we assign similarly a weight $w(v)$.
We wite $\sum_{w\in\Gamma} w$ for the sum over all these edge and vertex weights.
Then, we define
\be \omega_{2n}(\Gamma):=-2n|\Gamma|+\sum_{w\in\Gamma} w.\ee
Here, a  grading $|\Gamma|$ is used which is provided by the number of independent cycles in a graph 
$\Gamma$, -its lowest Betti number-, and we hence write 
\be H=\underbrace{H^0}_{\mathbb{Q}\One}\oplus\underbrace{\left(\oplus_{j=1}^\infty H^j\right)}_{\mathrm{Aug}(H)}.\ee
So $H$ is reduced to scalars off the augmentation ideal $\mathrm{Aug}(H)$.
We let $\langle\Gamma\rangle$ be the linear span of generators.

We distinguish these Hopf algebras $H=H_{2n}$ by an even integer $0\leq 2n$, $n\in\mathbb{N}$. They are all based on the same set of generators, hence have an identical algebra structure. 
There are slight differences in their coalgebra structure though, as we give them a  coproduct depending on $2n$:
\be \Delta_{2n}(\Gamma):=\Gamma\otimes\One+\One\otimes\Gamma+
\sum_{\gamma=\prod_i\gamma_i\subset\Gamma, \omega_{2n}(\gamma_i)\leq 0}\gamma\otimes\Gamma/\gamma.\ee
The sum is over all disjoint unions of 1-PI subgraphs $\gamma_i$ such that for each $\gamma_i$, $\omega_{2n}(\gamma_i)\leq 0$.
In the limit $n\to\infty$, we hence obtain the core Hopf algebra $H_{\mathrm{core}}$ with coproduct
\be \Delta_{\mathrm{core}}(\Gamma)=\Gamma\otimes\One+\One\otimes\Gamma+
\sum_{\gamma=\prod_i\gamma_i\subset\Gamma}\gamma\otimes\Gamma/\gamma.\ee
We also use the reduced coproducts
\be \Delta^\prime_{2n}(\Gamma):=
\sum_{\gamma=\prod_i\gamma_i\subset\Gamma, \omega_{2n}(\gamma_i)\leq 0}\gamma\otimes\Gamma/\gamma.\ee

This gives us a tower of quotient Hopf algebras \cite{BlKr}
\be H_0\subset H_2\subset H_4\subset\cdots\subset H_{2n}\cdots \subset H_{\mathrm{core}}.\ee
In the following, we often omit the subscript ${}_{2n}$ as it is either clear which integer we speak about, or the statement holds for 
arbitrary $2n$ in an obvious manner.

Note that $H_0$ is the trivial Hopf algebra in which every graph is primitive. 
It is the free commutative and cocommutative bialgebra of polynomials in all its generators $\in\langle\Gamma\rangle$. Fittingly, 
its use in zero-dimensional field theory is an excellent tool to count graphs \cite{0dimFT}.

For any other such $H_{2n}\in\mathcal{H}$, $n<\infty$, we find that the Hopf algebra decomposes into a semi-direct product
\be H_{2n}=H_{2n}^{\mathrm{ren}}\times H_{2n}^{\mathrm{ab}},\ee
where $H_{2n}^{\mathrm{ren}}$ is generated by graphs $\Gamma$ such that $\omega_{2n}(\Gamma)\leq 0$ and 
$H_{2n}^{\mathrm{ab}}$ is the abelian factor generated by graphs such that $\omega_{2n}(\Gamma)>0$. See \cite{RHI}.

Let us explain the above tower a bit more.
The core Hopf algebra allows to shrink any 1-PI subgraph $\gamma_i$ to a point, and hence is built on 
graphs with internal vertices of arbitrary valence, coupling an arbitrary numbers of edges and all types of edges for which we had free covariances.
Again, locality and representation theory provide for such vertices Feynman rules as before,
which are in general a sum over all local operators which are in accordance with the quantum numbers of those covariances.
We can distinguish those operators by labeled vertices, which does not hinder us to set up the Hopf algebra as before.
For the core Hopf algebra, all primitives which we find in the linear span $\langle\Gamma\rangle$ have degree one,
\be \Delta^\prime_{\mathrm{core}}(\Gamma)\not= 0\Rightarrow |\Gamma|>1\ee

Note that for any chosen finite $2n$, the results can be very different. 
A renormalizable theory is distinguished by the fact that for some finite $n_0$,
\be \omega_{2n_0}(\Gamma)=\omega_{2n_0}(\Gamma^\prime),\;\forall \,
\Gamma,\Gamma^\prime\;\mathrm{with}\;\mathrm{res}(\Gamma)=\mathrm{res}(\Gamma^\prime).\ee 
Here, $\mathrm{res}(\Gamma)$ is the map which assigns to
a graph $\Gamma$ the vertex obtained by shrinking all internal edges to zero length. What remains is the external edges connected to the same point.
If the number of external edghes was greater than two, this gives us a vertex. If it was two, we identify those two connected edges to a single edge. 

If such a $n_0$ exists, we call $2n_0$ the critical dimension of the theory. Particle physics so far is concerned with theories critical at 
$n_0=2$, ie.\ in four dimensions of spacetime.

In such a case, all graphs with the same type of external edges evaluate to the same result under evaluation by $\omega_{2n_0}$.
$\omega_{2n_0}(\Gamma)$ then takes values $\in\{-r_0,\cdots,+\infty\}$, where $-r_0$ is the value achieved for vacuum graphs,
and we obtain arbitrary positive values on considering graphs with a sufficient number of external edges. 

For $n>n_0$, for any configuration
of external edges we find, at sufficiently high degree $|\Gamma|$, graphs such that $\omega_{2n}(\Gamma)\leq 0$. The theory becomes non-renormalizable.

If $n<n_0$, only a finite number of graphs fulfils $\omega_{2n}(\Gamma)\leq 0$ and the theory is super-renormalizable.

In any case, for a Hopf algebra $H_{2n}$, continuing our appeal to self-similarity, we consider graphs made out of vertices such that 
$\omega_{2n}(\Gamma)\leq 0$. This defines a Hopf algebra $H_{2n}^{\mathrm{ren}}$.
Graphs made out of such vertices but with sufficiently many external edges such that $\omega_{2n}(\Gamma)>0$
then provide a semi-direct product $H_{2n}=H_{2n}^{\mathrm{ren}}\times H_{2n}^{\mathrm{ab}}$.
This Hopf algebra is a quotient of the core Hopf algebra, eliminating any graph with undesired vertices.

So already at this elementary level, there is a nice interplay between the before-mentioned representation theory of the Lorentz and Poincar\'e groups 
and such towers of Hopf algebras, as it is this representation theory which determines the covariances and their possible local vertices, and hence 
the quotient algebras we get. 

Let us now continue to list the other structural maps of those Hopf algebras. 
An antipode:
\be S(\Gamma)=-\Gamma-\sum_{\gamma\subset\Gamma}S(\gamma)\Gamma/\gamma.\ee
A counit $\bar{e}:H\to\mathbb{Q}$ and unit $E:\mathbb{Q}\to H^0\subset H$:
\be \bar{e}(q\One)=q, \bar{e}(X)=0, X\in \mathrm{Aug}(H),\;E(q)=q\One.\ee
Finally, an example:
\beas \Delta\left(\verttl\right) & = & 3\vertol\otimes\vertol\\  +2\feol\otimes\vertol+\phol\otimes\vertol. & & \eeas
As a final remark, note that there are many more quotient Hopf algebras, by restricting generators to planar, or parquet, or whatever graphs.

Also,  we will find all the Hopf algebras needed for an operator product expansion as quotient Hopf algebras, using that
for monomials (in operator-valued fields and their derivatives) $\mathcal{O}_a,\mathcal{O}_b$, the expansion of vacuum expectation values (vev's) 
of products of two monomials 
at different spacetime points $x,y$ into localized field monomials
\be \langle \mathcal{O}_a(x)\mathcal{O}_b(0)\prod_i\mathcal{O}_{d_i}(y_i)\rangle 
=
 \sum_{c} \mathcal{C}_{ab}^c(x)\,\langle\mathcal{O}_c(0) \prod_i\mathcal{O}_{d_i}(y_i)\rangle,\ee
(for $|x|<|y_i|\;\forall i$ and suitable (generalized) functions on spacetime $\mathcal{C}_{ab}^c$ determined only by the operators labelled $a,b,c$)
proceeds on a set of graphs having as local vertices the tree-level vev's of the operators $\mathcal{O}_c$, 
again in accordance with Wigner's representation theory. Note that all such vertices appear naturally in the core Hopf algebra (as we have  quotients
$\Gamma/\gamma$), and hence the core Hopf algebra is the endpoint in this tower of Hopf algebras which allows to formulate a full field algebra
in the sense of operator product expansions. In passing, we mention that the operator product expansion has a connection to vertex algebras as recently established by Hollands and Olbermann
\cite{HO}. 

Such expansions in the core Hopf algebra also then underly any study of the renormalization group flow in the sense of Wilson 
from the Hopf algebraic viewpoint. Let us finish this section with a cautionary remark: the difference between Minkowski or Euclidean signature 
is rather irrelevant for most combinatorial considerations below. It is crucial though in the operator product expansions, where the set of operators 
$\mathcal{O}_c$ above needs much more careful consideration in the Minkowskian case for exapnsions on the lightcone.

\subsection{The Lie algebra $L$ such that $U^*(L)=H$}
As a graded commutative Hopf algebra (\cite{Manchon}), any $H\in\mathcal{H}$ can be regarded as the dual $U^*(L)$ of the universal enveloping 
algebra $U(L)$ of a Lie algebra  $L$. The tower $\mathcal{H}$ of quotient Hopf algebras $H_{2n}$ 
corresponds to a tower $\mathcal{L}$ of sub-Lie algebras $L_{2n}$.
We write for each $L\in\mathcal{L}$,
\be U(L)=Q\One\oplus L\oplus_{k=2}^\infty L^{\hat{\otimes}^k},\ee
where $L^{\hat{\otimes}^k}$ indicates the symmetrized $k$-fold tensor product of $L$ as usual for an universal enveloping algebra,
obtained  by dividing the tensor algebra $L^{\otimes^k}$ by the ideal
$l_1\otimes l_2-l_2\otimes l_1-[l_1,l_2]=0$.

We manifest the duality by a pairing between generators of $L$ and generators of $H$,
\be (Z_\gamma,\Gamma)=\delta_{\gamma,\Gamma},\ee the Kronecker pairing.
It extends to $U(L)$ thanks to the coproduct.

There is an underlying pre-Lie algebra structure:
\be [Z_{\Gamma_1},Z_{\Gamma_2}]=Z_{\Gamma_1}\otimes Z_{\Gamma_2}-Z_{\Gamma_2}\otimes Z_{\Gamma_1}\ee
with 
\be [Z_{\Gamma_1},Z_{\Gamma_2}]=Z_{\Gamma_2\star\Gamma_1-\Gamma_1\star\Gamma_2}.\ee
Here, $\Gamma_i\star\Gamma_j$ sums over all ways of gluing $\Gamma_j$ into $\Gamma_i$,
which can be written as
\be \Gamma_i\star\Gamma_j=\sum_\Gamma n(\Gamma_i,\Gamma_j,\Gamma)\Gamma.\ee
For any $\Gamma\in H$, we have
\be ([Z_{\Gamma_1},Z_{\Gamma_2}],\Gamma)=(Z_{\Gamma_1}\otimes Z_{\Gamma_2}-Z_{\Gamma_2}\otimes Z_{\Gamma_1},\Delta(\Gamma)),\ee
for consistency.

With such section coefficients $n(\Gamma_i,\Gamma_j,\Gamma)$ we then have
\be \Delta(\Gamma)=\sum_{h,g}n(h,g,\Gamma)g\otimes h.\ee
The (necessarily finite, as $\Delta$ respects the grading) sum is over all graphs $h$ including the empty graph and all monomials in graphs $g$.

Note that we can regard  a graph $\Gamma$ obtained by inserting $\Gamma_j$ into $\Gamma_i$ as an extension
\be 0\to\Gamma_j\to\Gamma\to\Gamma_i\to 0.\ee
A proper mathematics discussion of this idea has been given recently by Kobi Kremnizer and Matt Szczesny \cite{KrMatt}.
\subsection{Hochschild cohomology}

The Hochschild cohomology is encaptured by non-trivial one-cocycles $B_+^\gamma:H\to \mathrm{Aug}(H)$.
The one cocycle condition (see \cite{BergbKr}) means \be bB_+^\gamma=0 \Leftrightarrow \Delta B_+^\gamma(X)=B_+^\gamma(X)\otimes 
\One+(\mathrm{id}\otimes B_+^\gamma)\Delta(X).\ee 
We define $\forall \gamma\in\langle\Gamma\rangle$, such that  $\Delta^\prime(\gamma)=0$, linear maps
\be B_+^\gamma(X):= \sum_{\Gamma\in \langle\Gamma\rangle}\frac{{\bf bij}(\gamma,X,\Gamma)}{|X|_\vee}\frac{1}{\mathrm{maxf}(\Gamma)}
\frac{1}{(\gamma|X)}\Gamma.\ee
Here, the sum is over the linear span $\langle\Gamma\rangle$ of generators of $H$.
Furthermore,\\
i)  $\mathrm{maxf}(\Gamma)$ is the number of maximal forests 
of $\Gamma$ defined as the integer 
\be \mathrm{maxf}(\Gamma)= \sum_{p,\gamma\in\langle \Gamma\rangle, \Delta^\prime(p)=0} (Z_\gamma,\Gamma^\prime)(Z_p,\Gamma^{\prime\prime}),\ee
(we used Sweedler's notation $\Delta(\Gamma)=\Gamma^\prime\otimes\Gamma^{\prime\prime})$\\
ii) $|X|_\vee$ is the number of distinct graphs obtained by permuting external edges of a graph,\\ iii) ${\bf bij}(\gamma,X,\Gamma)$ 
is the number of bijections between
the external edges of $X$ and half-edges of $\gamma$ such  that $\Gamma$ results,\\ 
iv) and finally $(\gamma|X)$ is the number  of insertion places for $X$ in $\gamma$.\\ 
Finally, for any $r$ which can appear as a residue $\mathbf{res}(\Gamma)$, we define
\be B_+^{r;k}=\sum_{\mathbf{res}(\gamma)=r,|\gamma|=k}\frac{1}{\mathrm{Aut}(\gamma)}B_+^\gamma,\ee
which sums over all $B_+^\gamma$ with a specified external leg structure and loop number, weighted by the rank $\mathrm{Aut}(\gamma)$ 
of their automorphism group.

\medskip

We want to understand these notions. 
We will do so by going through an example (see \cite{anatomy} for a more thorough exploration): 
\be \Gamma=\qedtlone+\qedtltwo+\qedtlthree+\qedtlfour.\ee
We will investigate
\be B_+^{\phol}\left( \vertol \feol \right)=? \label{bpluso}\ee
and 
\be B_+^{\phol}\left( \qedtwlone+\qedtwltwo \right)=? \label{bplust}\ee
Let us start with (\ref{bpluso}).
We have
\be \left| \vertol\feol  \right|_\vee=1.\ee
As fermion lines are oriented and hence all external edges distinguished, 
we can not permute external edges and obtain a different graph contributing to the same amplitude.
Now let us count the bijections.
\be {\bf bij}\left(\phol,\vertol\feol, X \right)  =  1,\ee
for all
$$X\in\left\{ \qedtlone,\qedtltwo,\qedtlthree,\qedtlfour\right\}.$$
Indeed, to glue the argument $X$ of $B_+^\gamma(X)$ into $\gamma$, we identify the factors $X=\prod_i\gamma_i$. 
The multiset $\mathrm{res}(\gamma_i)$
identifies a number of edges and vertices. 
>From the internal edges and vertices of $\gamma$ we choose a corresponding set $m$ which contains 
the same type and number of internal edges and vertices. 

We then consider the external edges of elements $\gamma_i$ of $X$  
and count bijections between this set and the similar set defined from $m$. 
Summing over all choices of $m$ and counting all bijections at a given $m$ 
such that $\Gamma$  is obtained gives ${\bf bij}$ by definition.
In the example, there is just a unique such bijection for each of the four different graphs $X$.

\be \left( \phol|\vertol\feol  \right)=4.\ee
This counts the number of insertion places. $\phol$ has two internal vertices and two internal edges, 
hence four possible choices of an insertion place.

Next, the maximal forests: we count the number of different subsets $\gamma$ of 1PI subgraphs such that $\Gamma/\gamma$ is a primitive element,
$\Delta^\prime(\Gamma/\gamma)=0$.

\be \mathrm{maxf}\left( X\right)=\mathrm{maxf}\left( \qedtltwo\right)= 2,\ee
for any of the four graphs $X$ as above.
For each of the four graphs there are two such possibilities. We indicate them in a way which makes the 
underlying tree structure (\cite{K,CK}) obvious:
\be \qedtlonetree.\ee
This is one major asset of systematically building graphs from images of Hochschild closed one-cocycles: it resolves for us overlapping divergences into 
rooted trees. 

Let us now collect:
\bea B_+^{\phol}\left( \vertol\feol\right)  =  & & \\
= \frac{1}{8}\left( \qedtlone+\qedtltwo+\qedtlthree+\qedtlfour \right). & &  \eea
The reader will notice that this fails to satisfy the desired cocycle property. 
To understand the reason for this failure and the solution
to this problem, we turn to  (\ref{bplust}).
We have
\be \left| \qedtwlone  \right|_\vee=\left| \qedtwltwo  \right|_\vee=1,\ee
as before.

\bea {\bf bij}\left(\phol,\qedtwlone,X\right) & = & 1,\\
     {\bf bij}\left(\phol,\qedtwltwo,X\right) & = & 1,\eea
where $X$ can still be any of the four graphs defined above.

Next,
\be \left( \phol|\qedtwlone\right)=2=\left( \phol|\qedtwltwo  \right).\ee
There are now two insertion places for the vertex graph to be inserted into the one-loop photon self-energy graph.

The maximal forests remain unchanged as we are generating the same graphs $X$ in the  two examples.
Hence
\bea B_+^{\phol}\left( \qedtwlone+\qedtwltwo\right) = & &\\
= \frac{1}{4}\left( \qedtlone+\qedtltwo+\qedtlthree+\qedtlfour \right). & &  \eea
Again, this fails to satisfy the cocycle property.
But let us now consider 
\bea B_+^{\phol}\left( 4 \vertol \feol +2\left( \qedtwlone+\qedtwltwo \right)\right) & &\\
=\left( \qedtlone+\qedtltwo+\qedtlthree+\qedtlfour \right). & &\label{multiplicity}
\eea
We see that with these weights we do fulfill the cocycle condition. For this, it is actually sufficient that the ratio of the weights is two-to-one.
Taking those weights to be four and two gives the result with the proper weights needed in the perturbative expansion of the photon propagator.
It was a major result of \cite{anatomy} that these weights always work out in field theory such that we do have the desired perturbative expansion
and cocycle properties. So while the maps $B_+^\gamma$ are one -cocycles for Hopf algebras generated by dedicated subsets of graphs, one finds that
the maps $B_+^{r;k}$ are proper cocycles for a Hopf algebra generated by sums of graphs with given external leg structure and loop number.

So working out 
\be B_+^{\phol}\left( \vertol \right)=\frac{1}{2}\qedtl,\ee
and
\be B_+^{\phol}\left( \feol \right)=\frac{1}{2}\qedtlf,\ee
we indeed confirm 
\be \Delta B_+^{\phol}(X)=B_+^{\phol}(X)\otimes\One+\left(\mathrm{id}\otimes B_+^{\phol}\right)\Delta(X),\ee
for 
\be X=4 \vertol \feol +2\left(\qedtlone+\qedtwltwo \right).\ee
We will understand soon how the weights $4$ and $2$ in (\ref{multiplicity}) come about. 

As a final exercise the reader might finally wish to confirm
\bea B_+^{\phol}\left( 2\vertol+2\feol \right) & = & \phtl,\\ 
\Delta B_+^{\phol}\left(2\vertol+2\feol\right) & = & B_+^{\phol}\left(2\vertol+2\feol\right)\otimes \One\nonumber\\
 & + &  \left( \mathrm{id} \otimes B_+^{\phol}  \right) \Delta \left( 2\vertol+2\feol\right).\nonumber\eea
To put it shortly: 
\be B_+^{\frac{1}{4}F^2;1}(2c_1^{\bar{\psi}A\!\!\!/\psi}+2c_1{\bar{\psi}\psi})=c_2^{\bar{\psi}A\!\!\!/\psi},\ee
where we indicated the residue $\mathbf{res}(\phol)$ of the one-loop primitive graph $\phol$ by 
its corresponding monomial $\frac{1}{4}F^2$ in the Lagrangian of QED, and there is indeed only one primitive at first loop order,
\be B_+^{\frac{1}{4}F^2;1}=B_+^{\phol}.\ee
\subsection{Sub-Hopf algebras}
In the example above, we looked at the sum of all 1-PI graphs contributing to a choosen amplitude $r$ at a given loop order $k$ .
This gives us Hopf algebra elements $c_k^r\in H^k$ as particular linear combinations of degree-homogenous elements.
Such Hopf algebra elements generate a sub-Hopf algebra. For example in QED we have 
\bea
\Delta^\prime(c_k^{\bar{\psi}A\!\!\!/\psi}) & = & \sum_{j=1}^{k-1} \left[ (2(k-j)+1)c_j^{\bar{\psi}A\!\!\!/\psi}
+2(k-j)c_j^{\bar{\psi}\psi}+(k-j) c_j^{\frac{1}{4}F^2}\right]\\ & &  
\otimes c_{k-j}^{\bar{\psi}A\!\!\!/\psi} +\mathrm{terms\;non-linear\;on\; the\;lhs}\nonumber\\
\Delta^\prime(c_k^{\bar{\psi}\psi}) & = & \sum_{j=1}^{k-1} \left[(2(k-j)) c_j^{\bar{\psi}A\!\!\!/\psi}+
(2(k-j)-1)c_j^{\bar{\psi}\psi}+(k-j)c_j^{\frac{1}{4}F^2}\right]\\ & & 
\otimes c_{k-j}^{\bar{\psi}\psi} +\mathrm{terms\;non-linear\;on\; the\;lhs}\nonumber\\
\Delta^\prime(c_k^{\frac{1}{4}F^2}) & = & \sum_{j=1}^{k-1} \left[(2(k-j)) c_j^{\bar{\psi}A\!\!\!/\psi}
+(2(k-j)-1)c_j^{\bar{\psi}\psi}+(k-j)c_j^{\frac{1}{4}F^2}\right] \\ & & 
\otimes c_{k-j}^{\frac{1}{4}F^2} +\mathrm{terms\;non-linear\;on\; the\;lhs}\nonumber.\eea
We omit to give explicit expressions for the terms non-linear on the lhs of the coproduct. They are not really needed, as we will soon see
when we study the Dynkin operator. Similar to these sub-Hopf algebras, one can determine the corresponding quotient Lie algebras.
\subsection{Co-ideals}
Often, sub-Hopf algebras like above only emerge when divided by suitable co-ideals.
An immediate application is a derivation of Ward--Takahashi and Slavnov--Taylor identities in this context \cite{anatomy,Walter}.
Lifting the idea of capturing relations between Green functions to the core Hopf algebra leads to the celebrated BCFW recursion relations \cite{WvSDK}.
All this needs much further work. The upshot is that dividing by a suitable co-ideal $I$, Feynman rules $\Phi:H\to \mathbb{C}$ 
can be well-formulated as maps
\be \Phi: H/I\to \mathbb{C}.\ee

Let us consider as an example (following van Suijlekom \cite{QEDWalter}) the ideal and co-ideal $I$ in QED given by 
\be i_k:=c_k^{\bar{\psi}A\!\!\!/\psi}+c_k^{\bar{\psi}\psi}=0,\,\forall k>0.\ee
So, for example,
\be i_1=\vertol+\feol.\ee
For $I$ to be a co-ideal we need  
\be \Delta(I)\subset (H\otimes I)\oplus (I\otimes H).\ee
Let us look at $\Delta(i_2)$ for an example:
\beas & & \Delta^\prime\left( \underbrace{\verttl+\fetl}_{\in I}\right) =  \\
 & & +\underbrace{ \phol \otimes \left( \vertol+\feol\right) }_{\in H\otimes I}  \\
 & & +\underbrace{ \left( \vertol+\feol \right) \otimes \left(2\vertol+\feol\right)}_{\in I\otimes H} \\
 & & +\underbrace{\vertol \otimes \left( \vertol+\feol \right) }_{\in H\otimes I}. 
\eeas
For a thorough discussion of the role of co-ideals and their interplay with Hochschild cohomology 
in renormalization and core Hopf algebras, see \cite{WvSDK} and references there.
\subsection{Co-radical filtration and the Dynkin operator}
For our graded commutative Hopf algebras $H$ there is a co-radical filtration. 
We consider iterations $[\Delta^\prime]^k: H\to \mathrm{Aug}(H)^{\otimes (k+1)}$ of the map 
$\Delta^\prime: H\to \mathrm{Aug}(H)\otimes \mathrm{Aug}(H)$, and filter Hopf algebra elements by the smallest integer $k$ such that
they lie in the kernel of such a map. 
We can write the Hopf algebra as a direct sum over the corresponding graded spaces $H^{[j]}$, \be H=\oplus_{j=0}^\infty H^{[j]}.\ee
Elements $q\One$ are in $H^{[0]}$, primitive elements are in $H^{[1]}$, and so on.

A Hochschild one-cocycle is now a map 
\be B_+^\gamma: H^{[j]}\to H^{[j+1]}.\ee
Note that for example in $H^{[2]}$,
\be B_+^x(\One) B_+^y(\One)= B_+^x\circ B_+^y(\One)+B_+^y\circ B_+^x(\One),\ee
with the difference between the lhs and the rhs being an element in $H^{[1]}$.

In \cite{BlKr} this was used to reduce the study of renormalization theory to the study of flags of subdivergent sectors.
This is closely connected to the Dynkin operator \cite{BroadK,KY2,Patras}
\be D:H\to \langle\Gamma\rangle,\;D:=S\star Y=m(S\otimes Y)\Delta.\ee
Here, $Y(\Gamma)=|\Gamma|\Gamma$ for all homogenous elements, extended by linearity.

Indeed, the above difference can be calculated as
\bea D(B_+^x\circ B_+^y(\One)+B_+^y\circ B_+^x(\One)) & = & (|x|+|y|)
\left(B_+^x\circ B_+^y(\One)+B_+^y\circ B_+^x(\One)\right.\nonumber\\ & & \left.-B_+^x(\One) B_+^y(\One)\right).\eea
In physics, this leads to the next-to-leading log expansion, see \cite{BroadK}, upon recognising that the Feynman rules send elements 
in $H$ to polynomials in suitable variables $L=\ln q^2/\mu^2$ say such that elements in $H^{[k]}$ are mapped to the terms $\sim L^k$.

There is an interesting remark to be made concerning the fact that the Dynkin operator vanishes on products.
This allows for all things concerning renormalization (including for example the derivation of the renormalization group \cite{RHII})
to rely on a linearized coproduct
\be \Delta_{\mathrm{lin}}:=(P_{\mathrm{lin}}\otimes \mathrm{id})\Delta: H\to H\otimes H,\ee
with $P_{\mathrm{lin}}:H\to \langle\Gamma\rangle$ the projector into the linear span of generators.

Obviously, this is not a coassociative map.
\be (\Delta_{\mathrm{lin}}\otimes\mathrm{id})\Delta_{\mathrm{lin}} \not= (\mathrm{id}\otimes \Delta_{\mathrm{lin}})\Delta_{\mathrm{lin}}.\ee
To control this loss of associativity is a fascinating task on which we hope to report in the future.
\subsection{Unitarity of the $S$-matrix}
A fact which will need much more attention in the future from the viewpoint of mixed Hodge structures is the fact that
Feynman amplitudes are boundary values of analytic functions. We hence have dispersion relations available, and can relate, in the spirit of the Cutkosky rules, 
branchcut ambiguities to cuts on diagrams.

In particular, following guidance of the core Hopf algebra whose primitives are the one-loop cycles in the graph, the structure of the following matrix
should reveal the desired relation between Feynman amplitudes and (variations of) mixed Hodge structures.

Actually, let us study a simple example where the renormalization Hopf algebra suffices (as the extra co-graphs in the core Hopf algebra would all
be tadpoles \cite{KrCMI}):
\bea 
\feolol=B_+^{\feolup}(\feol).
\eea
Then, the two-particle cuts on $\Gamma:=\feolol$ are given by the two-particle cuts on the primtives appearing in the one-cocycles:
\be B_+^{\feolcm}\left( \feol \right) =  \feololcl+\feololcr.\ee
The whole imaginary part can be obtained from this plus the three-particle cut $\feololcm$.

This can be combined into a nice matrix $M^\Gamma$ which indeed suggests to study the connection to mixed Hodge structures more deeply.
\be 
M^{\feololsmall}:=
\left(\begin{matrix}
 \One & 0 & 0 \\
 \feol & \feolcm & 0 \\
\feolol & \feololcl+\feololcr & \feololcm  
\end{matrix}\right).
\ee
In each column we cut one loop at a time, such that suitable linear combinations of columns 
will express the branchcut ambiguities of the first column. 

We hope that such matrices come in handy in an attempt to deepen the connection between Hodge theory and quantum fields, which started with the study
of limiting mixed Hodge structures and renormalization in a recent collaboration between Spencer Bloch and the author \cite{BlKr}.
While there it was the nilpotent orbit theorem which was at work in the back, we hope that the reader gets an idea from the above how we hope to farther
the connection to Hodge structures. This hopefully succeeds in giving a precise mathematical backbone to renormalizability and unitarity simultanously,
a feast notoriously missing in all attempts at quantum field theory (and gauge theories in particular) at present.
\subsection{Fix-point equations}
Let us finish this paper by listing the final fix-point equations (we give them for QED, and refer the reader to \cite{anatomy,Walter,WvSDK} for the general case)
which generate the whole Feynman graph expansion of QED. We discriminate between the two formfactors of the massive fermion, $m\bar{\psi}\psi$ 
for its mass and 
$\bar{\psi}\partial\!\!\!/\psi$ for its wave function renormalization. Let 
\be \mathcal{R}_{\mathrm{QED}}:=\{\bar{\psi}\partial\!\!\!/\psi,m\bar{\psi}\psi,\bar{\psi}A\!\!\!/\psi,\frac{1}{4}F^2\}.\ee
Then
\be
X^{r}(\alpha)  =  \One\pm
\sum_{k=1}^\infty\alpha^k B_+^{r;k}(X^{r}(\alpha)Q^{2k}(\alpha)),\ee
where we take the plus sign for $r=\bar{\psi}A\!\!\!/\psi$ and the minus sign else, if $r$ corresponds to an edge.
We let \be Q=\frac{X^{\bar{\psi}A\!\!\!/\psi}}{ X^{\bar{\psi}\partial\!\!\!/\psi}\sqrt{X^{\frac{1}{4}F^2}}}.\ee
Upon evaluation by renormalized Feynman rules it delivers the invariant charge of QED.
The resulting maps $B_+^{r;k}$ are Hochschild closed \be bB_+^{i,K}=0.\ee 
Dividing by the (co-)ideal $I$ simplifies $Q$;
\be Q=\frac{1}{\sqrt{X^{\frac{1}{4}F^2}}}.\ee
See for example \cite{KY3} for a far-reaching application of these techniques in QED.

Let us finally mention that upon adding suitable exact terms, $B_+^{r;k}\to B_+^{r;k}+L_0^{r;k}$ with $ L_0^{r;k}=b\phi^{r,k}$, 
$b$ being the Hochschild differential $b^2=0$,
$\phi^{r,k}:H\to \mathbb{C}$, we can capture the change of parameters in the Feynman rules by suitable such coboundaries \cite{Kreimerprep}.

\end{document}